\documentclass[journal=jacsat,manuscript=article]{achemso}
\setkeys{acs}{articletitle = true}

\usepackage[version=3]{mhchem} 
\usepackage[T1]{fontenc}       
\usepackage{amssymb}           

\author{Man Yan Eric Yau}
\affiliation[OS]{Institut f\"ur Chemie neuer Materialien, Universit\"at Osnabr\"uck, Barbarastr.7, 49076 Osnabr\"uck, Germany}
\author{Ilja Gunkel}
\affiliation[MLU]{Martin-Luther-Universit{\"a}t Halle-Wittenberg, Institut f{\"u}r Physik, D-06099 Halle, Germany}
\alsoaffiliation{Current address: Adolphe Merkle Institute, Chemin des Verdiers 4, 1700 Fribourg, Switzerland}
\author{Brigitte Hartmann-Azanza}
\affiliation[OS]{Institut f\"ur Chemie neuer Materialien, Universit\"at Osnabr\"uck, Barbarastr.7, 49076 Osnabr\"uck, Germany}
\author{Wajiha Akram}
\affiliation[OS]{Institut f\"ur Chemie neuer Materialien, Universit\"at Osnabr\"uck, Barbarastr.7, 49076 Osnabr\"uck, Germany}
\author{Yong Wang}
\affiliation[NT]{Nanjing Tech University, State Key Lab of Materials-Oriented Chemical Engineering; College of Chemical Engineering, Xin Mofan Road 5, Nanjing 210009, Jiangsu, China}
\author{Thomas Thurn-Albrecht}
\email{thomas.thurn-albrecht@physik.uni-halle.de}
\affiliation[MLU]{Martin-Luther-Universit{\"a}t Halle-Wittenberg, Institut f{\"u}r Physik, D-06099 Halle, Germany}
\author{Martin Steinhart}
\email{martin.steinhart@uos.de}
\affiliation[OS]{Institut f\"ur Chemie neuer Materialien, Universit\"at Osnabr\"uck, Barbarastr.7, 49076 Osnabr\"uck, Germany}

\title[An \textsf{achemso} demo]
{Semicrystalline block copolymers in rigid confining nanopores}


\begin{document}




\begin{abstract}
We have investigated PLLA crystallization in lamellae-forming PS-\textit{b}-PLLA confined to straight cylindrical nanopores under weak confinement (nanopore diameter $D$ / equilibrium PS-\textit{b}-PLLA period $L_0 \geq$ 4.8). Molten PS-\textit{b}-PLLA predominantly forms concentric lamellae along the nanopores, but intertwined helices occur even for $D$/$L_0 \approx$ 7.3. Quenching PS-\textit{b}-PLLA melts below $T_G$(PS) results in PLLA cold crystallization strictly confined by the vitrified PS domains. Above $T_G$(PS), PLLA crystallization is templated by the PS-\textit{b}-PLLA melt domain structure in the nanopore centers, while adsorption on the nanopore walls stabilizes the outermost cylindrical PS-\textit{b}-PLLA shell. In between, the nanoscopic PS-\textit{b}-PLLA melt domain structure apparently ripens to reduce frustrations transmitted from the outermost immobilized PS-\textit{b}-PLLA layer. The onset of PLLA crystallization catalyzes the ripening while transient ripening states are arrested by advancing PLLA crystallization. Certain helical structure motifs persist PLLA crystallization even if PS is soft. The direction of fastest PLLA crystal growth is preferentially aligned with the nanopore axes to the same degree as for PLLA homopolymer, independent of whether PS is vitreous or soft.
\end{abstract}

\section{Introduction}

As compared to the bulk, both the crystallization of polymers and microphase separation of block copolymers (BCPs) are significantly modified under the two-dimensional confinement of rigid cylindrical nanopores. Homopolymer crystallization inside cylindrical nanopores \cite{Cryst-NP_Steinhart2003,Cryst-NP_Steinhart2006,Cryst-NP_Wu2007,Cryst-NP_Shin2007,Cryst-NP_Lin2012,Cryst-NP_Maiz2012,Cryst-NP_Maiz2013,Cryst-NP_Wu2013,Cryst-NP_Guan2013,Cryst-NP_Martin2013,Cryst-NP_Michell2014,PLLA_Guan2015,Cryst-NP_Michell2016} is typically dominated by kinetics. Homopolymer crystals having their directions of fast crystal growth aligned with the nanopore axes rapidly grow along the nanopores and prevail over crystals having other orientations. On the other hand, the nanoscopic morphologies formed by BCPs inside rigid cylindrical nanopores are dictated by BCP architecture, reduced nanopore diameter (ratio of nanopore diameter \textit{D} and bulk BCP period \textit{L}$_0$), interactions between the blocks and interactions with the nanopore walls. If \textit{D} is several times larger than \textit{L}$_0$, symmetric BCPs forming lamellae in the bulk often form ``dartboard'' morphologies -- concentric cylinder shells oriented parallel to the nanopore axes that alternatingly consist of the two BCP components \cite{BCP-NP_Shin2004,BCP-NP_Xiang2004,BCP-NP_Xiang2005b,BCP-NP_Sun2005,BCP-NP_Ma2009}. For \textit{D}/\textit{L}$_0$ $\lesssim$ 3 the lamellae were oriented normal to nanopore axes (``stacked-disks'') \cite{BCP-NP_Shin2004}. Simulations 
\cite{BCP-NP_He2001,BCP-NP_Sevink2001,BCP-NP_Feng2006a,BCP-NP_Feng2006c,BCP-NP_Li2006,BCP-NP_Wang2007,BCP-NP_Yu2007,BCP-NP_Sevink2008,BCP-NP_Li2009a,BCP-NP_Li2009b,BCP-NP_Pinna2009,BCP-NP_Huh2010} revealed that ``dartboard'' morphologies may occur for nanopore walls having preferential interactions with one block and ``stacked-disks'' morphologies if the interactions between nanopore walls and both blocks are balanced. As the BCP/nanopore wall interactions are tuned from selective to balanced, the transition between concentric-cylinder-shell morphology and stacked-disks morphology is characterized by complex structure motifs such as helices that have been interpreted either as equilibrium morphologies \cite{BCP-NP_Feng2006a,BCP-NP_Feng2006c,BCP-NP_Wang2007,BCP-NP_Yu2007,BCP-NP_Li2009a,BCP-NP_Li2009b,BCP-NP_Pinna2009} or as long--lived kinetically trapped states \cite{BCP-NP_Sevink2001,BCP-NP_Sevink2008}. 

Crystallization in bulk semicrystalline BCPs consisting of a crystallizable and an amorphous block may occur in the breakout mode, the templated mode and the confined mode \cite{Cryst-NP_Loo2002,Cryst-NP_He2012}. In the confined crystallization regime, crystallization is confined by the nanoscopic melt domain structure of the BCP that is typically fixated by vitrification of the non--crystallizing component \cite{Cryst-NP_Loo2001}. In the templated regime \cite{Cryst-NP_Loo2000} crystallization disturbs the nanoscopic BCP domain structure but does not completely destroy it; the nanoscopic domain structure is retained with local distortions and defects. Templated crystallization typically takes place in the presence of a soft matrix when crystallizable blocks and matrix are highly incompatible. In the breakout regime, crystallization destroys the nanoscopic BCP domain structure. 

Semicrystalline BCPs confined to rigid cylindrical nanopores crystallize under the hierarchical confinement imposed by the nanopore walls and the nanoscopic BCP morphology. Poly(ethylene) (PE) in PE-\textit{b}-PS (PS = polystyrene) inside cylindrical nanopores with $D \leq$ 60 nm crystallizes at high supercooling and melts with significant melting point depression \cite{Cryst-NP_Michell2012}. For asymmetric PS-\textit{b}-PE the PE blocks formed cylindrical shells surrounding PS cores. For symmetric PS-\textit{b}-PE an outer PE shell surrounding a PS core containing another, distorted PE domain running along the nanopore axes was found \cite{Cryst-NP_Casas2015}. However, crystallization of semicrystalline BCPs in cylindrical nanopores has attracted only limited interest. The interplay of the nanoscopic melt domain structures BCPs form under cylindrical confinement and crystallization is hardly understood, especially for weak confinement where $D$ is a few times larger than $L_0$. Here we study the crystallization of the poly-L-lactide (PLLA) blocks of lamellae--forming PS-\textit{b}-PLLA (PLLA volume fraction 50 \%; PS is atactic) located in the cylindrical nanopores of self--ordered anodic aluminum oxide (AAO) \cite{Cryst-NP_Masuda1998} under weak confinement ($D$/$L_0 \geq 4.8$). We crystallized the PLLA blocks either by cooling from the melt at --1 K/min or isothermally at crystallization temperatures $T_C$ = 140$^\circ$C (PS is soft) and $T_C$ = 85$^\circ$C (PS is vitreous). PS-\textit{b}-PLLA was selected as model BCP because of the strong segregation between the PS and PLLA blocks \cite{PLLA_Zalusky2002,PLLA_Ho2009a}. The Flory-Huggins parameter $\chi$($T$) of PS and PLLA equals 154.9/$T$ -- 0.211 \cite{PLLA_Ho2009a}; at 298 K $\chi$ amounts to 0.3 so that the PS-\textit{b}-PLLA used here is far in the strong segregation regime. Furthermore, poly(lactide) can be degraded selective to PS \cite{PLLA_Zalusky2001,PLLA_Zalusky2002}. Thus, PS nanorods containing voids in place of the PLLA domains are accessible, the internal morphology of which can be imaged by transmission electron microscopy (TEM). 

\section{Experimental Section}\label{exp_sec}

\subsection{Polymers}
PS-\textit{b}-PLLA (50 vol--\%-PLLA; $M_n$ (PS) = 21000 g/mol; $M_n$ (PLLA) = 24300 g/mol; $M_w$/$M_n$ 1.14) and PLLA homopolymer ($M_w$ = 16500 g/mol; $M_n$ = 13500 g/mol; $M_w$/$M_n$ = 1.20) were obtained from Polymer Source Inc., Dorval, Canada. Differential scanning calorimetry (DSC) revealed that quenching bulk PS-\textit{b}-PLLA at a rate of --160 K/min suppresses crystallization completely (Supporting Text 1; Supporting Figure S1). SAXS patterns of bulk PS-\textit{b}-PLLA quenched from the melt (Supporting Text 2; Supporting Figure S3a) indicate that an ordered lamellar domain structure with a spacing \textit{L}$_0$ = 37 nm formed without interference by PLLA crystallization, that is, the bulk PS-\textit{b}-PLLA melt morphology persisted thermal quenching. Cooling PS-\textit{b}-PLLA at --1 K/min (Figure S3b) from the molten state yielded SAXS patterns that do not show any significant features indicating the presence of a well-ordered nanoscopic domain structure (Supporting Text 2; Supporting Figure S3b). 

\subsection{Templates}
Self-ordered nanoporous anodic aluminum oxide (AAO) with a pore diameter \textit{D} of 180 nm, a pore depth of 60 $\mu$m and a nearest-neighbor distance of 500 nm was prepared following procedures reported previously \cite{Cryst-NP_Masuda1998}. The self--ordered AAO membranes used to prepare the PS-\textit{b}-PLLA nanorods shown in Figure \ref{TEM_85}d and Figure \ref{TEM_310nm} were prepared exactly in the same way except that the AAO nanopores were widened to \textit{D} = 275 nm -- 310 nm by isotropic etching with 10 wt-\% aqueous phosphoric acid at 30$^\circ$C. The self-ordered AAO membranes contained arrays of separated and aligned cylindrical nanopores with closed nanopore bottoms and were connected to a $\sim$ 940 $\mu$m thick supporting aluminum substrate.  
 
\subsection{Infiltration of self-ordered AAO}\label{methods:infiltration}
PS-\textit{b}-PLLA and PLLA homopolymer were located on the surfaces of self--ordered AAO. The polymer--covered self-ordered AAO membranes were heated to 180$^\circ$C at a rate of 10 K/min, kept at this temperature for 48 h and quenched with liquid nitrogen. Residual polymer was removed from the surface of the self-ordered AAO membranes using sharp blades. The quenched samples were again heated to 180$^\circ$C at a rate of 10 K/min and kept at this temperature for 12 h. For isothermal crystallization, the temperature was quenched to the crystallization temperature $T_C$ and kept at $T_C$ for 24 h. Non-isothermal crystallization was carried out by cooling the samples from 180$^\circ$C to room temperature at --1 K/min. No bulk polymer remained on the AAO surface during crystallization. Hence, the AAO nanopores contained separated polymer nanorods so that crystallization had to be initiated separately in each AAO nanopore. All high-temperature steps were carried out under argon atmosphere. 

\subsection{Wide-angle X-ray scattering (WAXS)}
\begin{figure}[H]
\centerline{\includegraphics[scale=0.6]{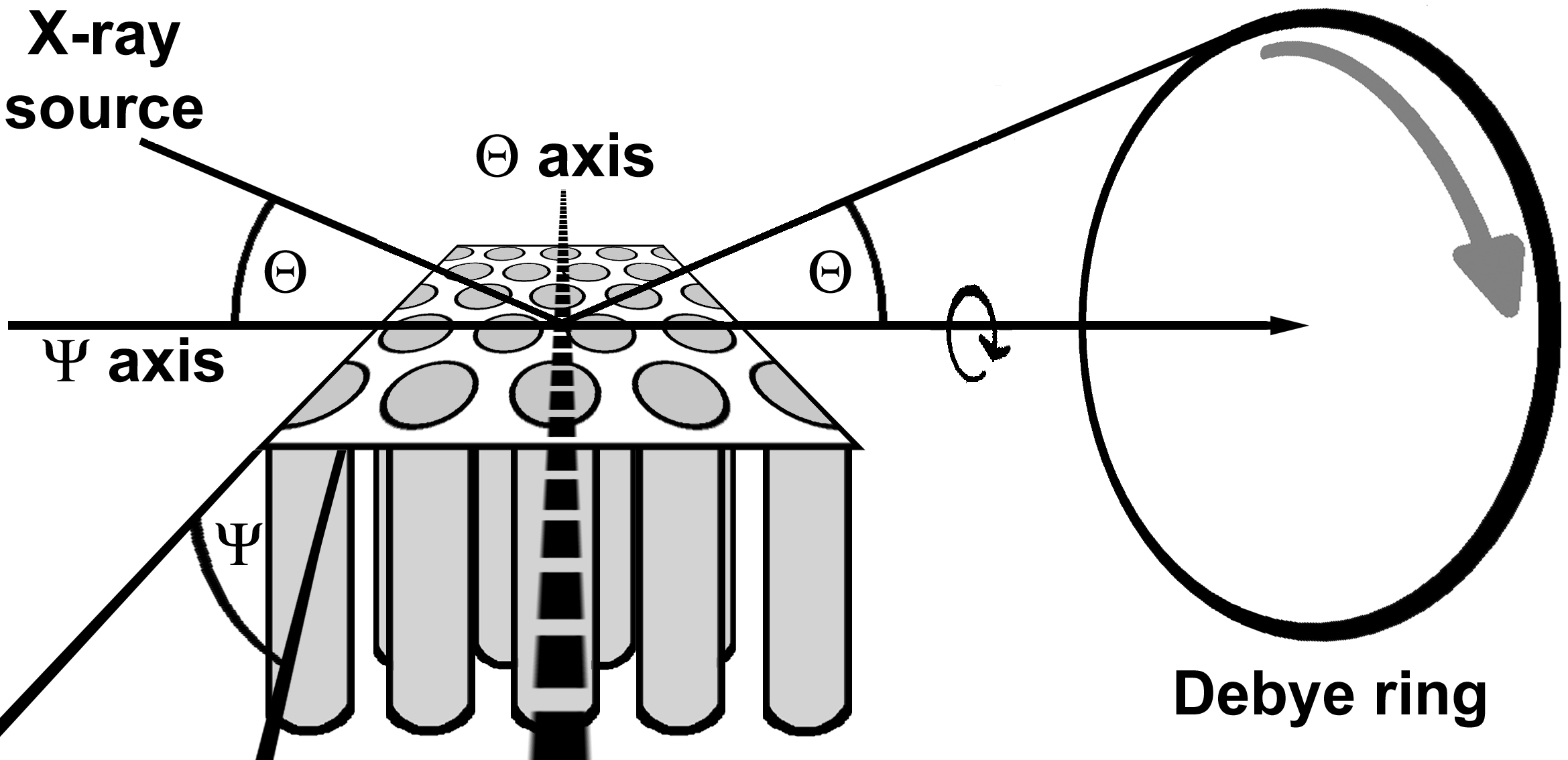}}
	\caption{Scattering geometry used for WAXS measurements on either PS-\textit{b}-PLLA or PLLA homopolymer located inside aligned cylindrical nanopores of self--ordered AAO. Samples can be rotated about the $\Theta$ axis by the scattering angle $\Theta$ and about the $\Psi$ axis by an azimuthal angle $\Psi$. $\Theta$/2$\Theta$ scans are measured by rotating the sample about the $\Theta$ axis. The scan follows a radial trajectory starting at the center of the reciprocal space while $\Psi$ is fixed. In the course of a Schulz scan, the sample is tilted about the $\Psi$ axis at fixed scattering angle $\Theta$ so that the scan follows the Debye-Scherrer rings corresponding to the selected scattering angle $\Theta$.}
\label{Schulz}
\end{figure}

For WAXS measurements on PS-\textit{b}-PLLA and PLLA homopolymer confined to self-ordered AAO membranes still attached to aluminum substrates we used a PANalytical X'Pert Pro MRD diffractometer operated with Cu $K\alpha$ radiation ($\lambda$ = 1.54 nm) configured for texture analysis of thin samples. The samples were mounted onto a Eulerian cradle that could be rotated about three axes. A proportional counter was used as point detector in combination with a Soller slit located between point detector and sample. While this configuration enables full mapping of the reciprocal space, only scattering intensity originating from the Bragg reflection under investigation is detected, while all other radiation is filtered out. The samples investigated here contain ensembles of aligned cylindrical AAO nanopores with uniaxial symmetry. Since the orientations of the PLLA crystals in the different AAO nanopores are not correlated, it is to be expected that powder--like properties are observed in any direction normal to the AAO nanopores. However, along the AAO nanopores anisotropic features should be detectable. The full range of relevant scattering patterns can be acquired by a combination of $\Theta$/2$\Theta$ scans and Schulz scans \cite{Cryst-NP_Schulz1949} (Figure \ref{Schulz}). Moreover, the sample volume probed by grazing--incidence techniques is typically restricted to thin layers close to the sample surface. The $\Theta$/2$\Theta$ scans and Schulz scans used here, however, probe the entire volume of the self--ordered AAO membranes so that the obtained diffraction patterns are representative of the entire sample volume.

During the $\Theta$/2$\Theta$ scans the AAO nanopores and the polymeric nanorods inside the AAO nanopores were oriented parallel to the scattering plane defined by the wave vector of the incident X-rays and the position of the point detector. The scattering angle $\Theta$ is the angle between the wave vector of the incident X-rays and the AAO membrane surface as well as the angle between the wave vector of the scattered X-rays pointing from the center of the Ewald sphere towards the point detector and the AAO membrane surface. During the $\Theta$/2$\Theta$ scans the X--ray source was fixed while the samples were tilted by an angle $\Theta$ about the $\Theta$ axis normal to the scattering plane as well as to the long axes of the AAO nanopores (Figure \ref{Schulz}). At the same time, the detector was moved along a circle centering about the $\Theta$ axis by an angle of 2$\Theta$. The second relevant angle $\Psi$ (cf. Figure \ref{Schulz}), which is the angle between AAO nanopore axes and scattering vector, was set to 0$^\circ$. In the course of a $\Theta$/2$\Theta$ scan, the length of the scattering vectors oriented parallel to the AAO nanopores was successively increased. Under these conditions, only scattering originating from sets of lattice planes oriented normal to the AAO nanopore axes and parallel to the AAO membrane surface contributed to the detected scattering intensity. 

Schulz scans \cite{Cryst-NP_Schulz1949} yield the scattering intensity \textit{I} at fixed $\Theta$ (sample) and 2$\Theta$ (detector) angles as function of $\Psi$. The obtained I($\Psi$) profiles correspond to the azimuthal scattering intensity profiles along the Debye-Scherrer rings of the Bragg reflections belonging to the fix $\Theta$ values. $\Psi$ is also the angle enclosed by a set of lattice planes with a specific \textit{d} value that meets the Bragg condition and the surface of the self--ordered AAO membrane. Since, if the Bragg condition is met, the reciprocal lattice vector belonging to the specific reflection under investigation and the scattering vector must coincide, $\Psi$ is then enclosed by the reciprocal lattice vector and the long axes of the AAO nanopores. Schulz scans are acquired  by tilting the sample about the $\Psi$ axis, which is the intersection of the surface of the self--ordered AAO membranes and the scattering plane (Figure \ref{Schulz}). Thus, $\Theta$ axis and $\Psi$ axis are oriented orthogonal with respect to each other. Note that for $\Psi$ angles larger than $\sim$70$^\circ$ defocussing effects occur \cite{Cryst-NP_Tenckhof1970}. 

\subsection{Transmission electron microscopy (TEM)}
For TEM investigations, PS-\textit{b}-PLLA nanorods were released from self-ordered AAO. At first, the aluminum substrate underneath the self-ordered AAO layer was etched with a solution of 1.7 g CuCl$_2$ $\cdot$ H$_2$O in 50 ml 37\% HCl and 50 ml deionized water at 0$^\circ$C. Then, the self-ordered AAO layer was etched with an aqueous 40 wt--\% KOH solution for a few minutes. This treatment also resulted in the degradation of the PLLA blocks by hydrolysis (cf. Supporting Text 3). Therefore, voids formed in place of the PLLA domains that can easily be identified in TEM images. The KOH solution was replaced by deionized water by several cycles including centrifugation, removal of the supernatant solution and redispersion of the precipitate in deionized water. After neutralization, some droplets of the nanorod suspensions were deposited onto copper grids coated with holey carbon films. TEM investigations were carried out with a JEOL 1010 microscope operated at 100 keV. 

\begin{figure}[H]
\centerline{\includegraphics[scale=0.2]{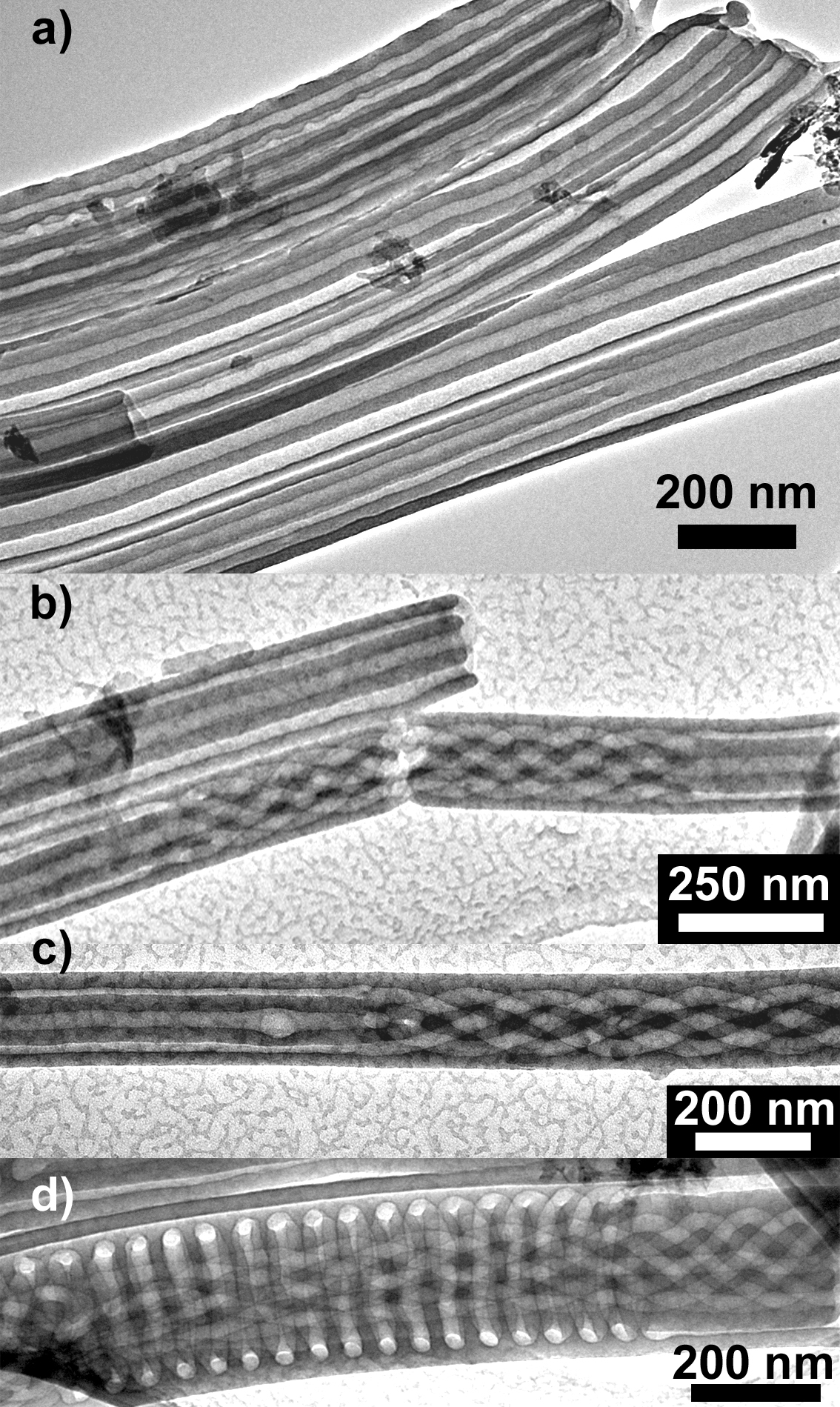}}
	\caption{TEM images of released PS-\textit{b}-PLLA nanorods after isothermal crystallization at $T_C$ = 85$^\circ$C. The PLLA initially located at the positions of the voids was hydrolyzed. a) Overview. b-d) Examples of PS nanorods containing hollow helices in place of the PLLA domains; b), c) \textit{D}/\textit{L}$_0$ $\approx$ 4.8; d) \textit{D}/\textit{L}$_0$ $\approx$ 7.3.}
\label{TEM_85}
\end{figure}

\section{Results}
\subsection {Morphology of PS-\textit{b}-PLLA nanorods}
Figure \ref{TEM_85}a-c shows PS nanorods obtained by infiltration of PS-\textit{b}-PLLA into self--ordered AAO (\textit{D}/\textit{L}$_0$ $\approx$ 4.8) followed by isothermal crystallization at $T_C$ = 85$^\circ$C (PS is glassy) and etching the AAO as well as the PLLA. The nanoscopic domain structure predominantly consists of alternating PS (conserved) and PLLA (etched) domains forming concentric cylinder shells (Figure \ref{TEM_85}a). Typically, the remaining PS nanorods consisted of two concentric PS cylinders, a central cylindrical void and a void between the concentric PS cylinders. Assuming the initial presence of an outermost PLLA shell, this morphology is in good agreement with a \textit{D}/\textit{L}$_0$ value of 4.8 (Figure \ref{TEM_85}b and c). Despite \textit{D}/\textit{L}$_0$ $\approx$ 4.8 the PS nanorods contain sections with non--classical confinement--induced morphologies, such as five intertwined hollow helices winding about a central straight cylindrical void with a lead (distance along intertwined helices covered by one complete rotation of a single helix) of $\approx$500 nm (Figure \ref{TEM_85}b) and four intertwined hollow helices with a lead of $\approx$600 nm (Figure \ref{TEM_85}c). Different morphologies may occur along one and the same PS-\textit{b}-PLLA nanorod. Examples for transitions between different morphologies within PS-\textit{b}-PLLA nanorods are shown in Figure \ref{TEM_85}b (transitions between concentric cylinder shells and five intertwined helices winding about a central straight cylinder) as well as in Figure \ref{TEM_85}c (transition between concentric cylinder shells and four intertwined helices). Even for \textit{D}/\textit{L}$_0$ $\approx$ 7.3 complex helical structure motifs were found, as shown in Figure \ref{TEM_85}d. An outermost single hollow helix with a pitch of $\approx$60 nm surrounds six intertwined inner hollow helices with a lead of $\approx$780 nm and a central straight cylindrical void. In the nanorod segment at the right side of Figure \ref{TEM_85}d, the morphology transforms into an outermost hollow cylinder shell surrounding five inner intertwined hollow helices with a lead of $\approx$380 nm and a central straight cylindrical void. 

\begin{figure}[H]
\centerline{\includegraphics[scale=0.2]{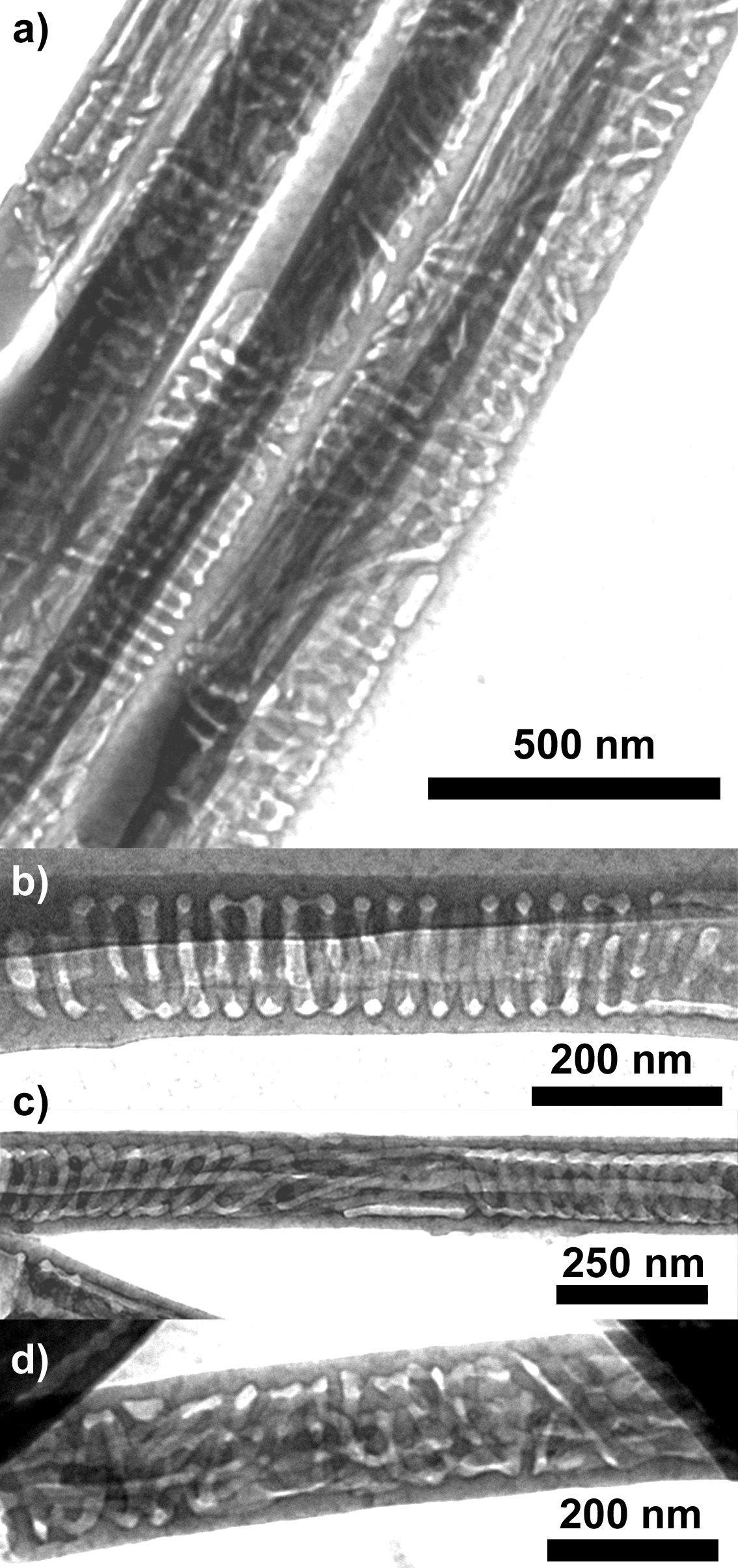}}
\caption{TEM images of released PS-\textit{b}-PLLA nanorods (\textit{D}/\textit{L}$_0$ $\approx$ 4.8) after isothermal crystallization at $T_C$ = 140$^\circ$C. The PLLA initially located at the positions of the voids was hydrolyzed. a) Overview. b)-d) Characteristic morphologies. b) Single hollow undistorted ``low--pitch'' helix with a pitch of $\approx$40 nm; c) distorted hollow helix surrounding a central hollow cylinder; d) disrupted concentric-lamellar morphology.}
\label{TEM_140}
\end{figure}

\begin{figure}[H]
\centerline{\includegraphics[scale=1]{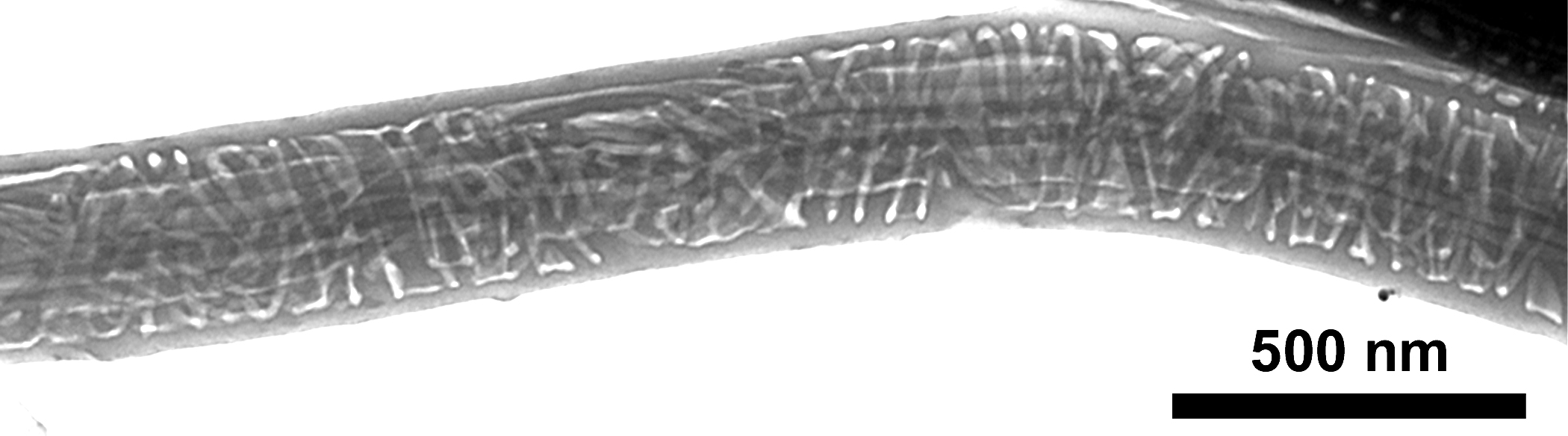}}
\caption{TEM image of a released PS-\textit{b}-PLLA nanorod (\textit{D}/\textit{L}$_0$ $\approx$ 8.4) after isothermal crystallization at $T_C$ = 140$^\circ$C in an AAO nanopore with a diameter of $\approx$310 nm. The PLLA initially located at the positions of the voids was hydrolyzed.}
\label{TEM_310nm}
\end{figure}

Figure \ref{TEM_140} shows PS nanorods (\textit{D}/\textit{L}$_0$ $\approx$ 4.8) obtained at $T_C$ = 140$^\circ$C (PS is soft) after etching the AAO and the PLLA blocks. Neither undisturbed concentric-cylinder-shell ``dartboard'' morphologies nor undisturbed intertwined multiple helices with leads of a few 100 nm were found (Figure \ref{TEM_140}a). Figure \ref{TEM_140}b--c shows three characteristic morphology types obtained at $T_C$ = 140$^\circ$C. Several PS nanorod segments contain undistorted single ``low-pitch'' helices -- the pitch of the helix seen in Figure \ref{TEM_140}b amounts to $\approx$40 nm approximately corresponding to \textit{L}$_0$. Moreover, distorted helices occurred, as displayed in Figure \ref{TEM_140}c. The most characteristic morphology type is shown in Figure \ref{TEM_140}d for \textit{D}/\textit{L}$_0$ $\approx$ 4.8 and in Figure \ref{TEM_310nm} for \textit{D}/\textit{L}$_0$ $\approx$ 8.4. The outermost PS cylinder shell remained mostly intact. However, the void next to the outermost PS cylinder shell -- after isothermal crystallization at $T_C$ = 85$^\circ$ typically intact -- was disrupted and often even dominated by PLLA domains oriented nearly normal to the AAO nanopore axes, such as distorted rings and helices. Cylinder shells closer to the center of the PS-\textit{b}-PLLA nanorods as well as their central cylindrical domains coinciding with the nanorod axes often remained mostly intact. All apparent morphological features identified after annealing at $T_C$ = 140$^\circ$C are characterized by length scales compatible with the bulk period $L_0$. 
 
\subsection {Orientation of PLLA crystals in PS-\textit{b}-PLLA confined to the cylindrical nanopores of self-ordered AAO}
\begin{figure}[H]
\centerline{\includegraphics[scale=0.7]{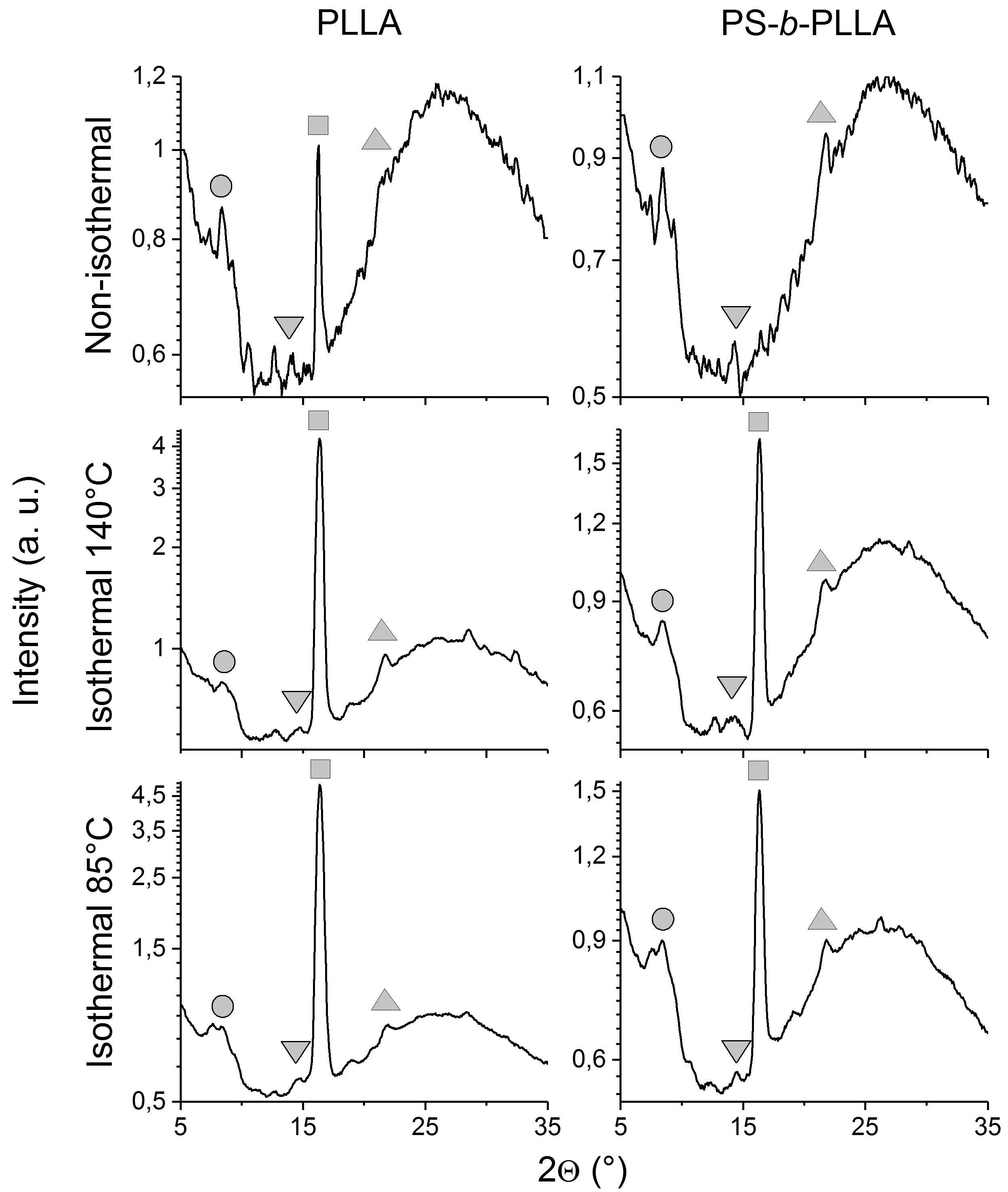}}
	\caption{Wide-angle X--ray patterns of PS-\textit{b}-PLLA and PLLA homopolymer located inside the parallel nanopores of self--ordered AAO (\textit{D} = 180 nm) after non-isothermal crystallization at --1 K/min and isothermal crystallization at $T_C$ = 140$^\circ$C or $T_C$ = 85$^\circ$C. In the $\Theta$/2$\Theta$ scattering geometry used here, sets of lattice planes oriented normal to the AAO nanopore axes and parallel to the AAO surface contributed to the detected scattering intensity. Circles denote (100) reflections, downward-triangles (010) reflections, squares (110)/(200) composite reflections and upward-triangles (210) reflections of orthorhombic pseudo-hexagonal $\alpha$--PLLA.}
	\label{WAXS}
\end{figure}

We were not able to acquire meaningful DSC thermograms of PS-\textit{b}-PLLA confined to self--ordered AAO (cf. Supporting Text 3). However, we could investigate PS-\textit{b}-PLLA and, for comparison, PLLA homopolymer crystallized inside self--ordered AAO with a nanopore diameter of 180 nm by WAXS using a diffractometer with a configuration specifically optimized for texture analysis of thin samples. The obtained $\Theta$/2$\Theta$ patterns (Figure \ref{WAXS}) indicate the presence of orthorhombic pseudo-hexagonal $\alpha$-PLLA \cite{PLLA_Miyata1997,PLLA_Iwata1998}. The (100) reflection at 2$\Theta$ $\approx$ 8.5$^\circ$ marked by circles is relatively pronounced in the WAXS patterns of non--isothermally crystallized PLLA homopolymer and of non--isothermally crystallized PS-\textit{b}-PLLA. All WAXS patterns show weak (010) reflections at 2$\Theta$ $\approx$ 14.5$^\circ$ marked by downward-triangles. The (110)/(200) composite reflection was reported to be by far the most intense reflection of $\alpha$--PLLA \cite{PLLA_Kobayashi1995}. Marked by squares, the (110)/(200) composite reflection at 2$\Theta \approx$ 16.3$^\circ$ indeed dominates the WAXS patterns of all isothermally crystallized samples as well as the WAXS pattern of non--isothermally crystallized PLLA homopolymer. However, the (110)/(200) composite reflection is completely absent in the $\Theta$/2$\Theta$ pattern of non--isothermally crystallized PS-\textit{b}-PLLA. Moreover, for bulk PLLA a systematic shift of the position of the (110)/(200) composite reflection to higher 2$\Theta$ values with increasing $T_C$ was reported \cite{PLLA_Zhang2005,PLLA_Kawai2007}. For PLLA homopolymer and PS-\textit{b}-PLLA confined to AAO nanopores no such shift was observed; the peak position remained by and large unaltered independent of the applied thermal treatment. Another weak reflection appearing in all patterns at 2$\Theta$ $\approx$ 21.8$^\circ$ marked by upward-triangles can be indexed as the (210) reflection of $\alpha$--PLLA \cite{PLLA_Iwata1998}. The (203) reflection of $\alpha$--PLLA at 2$\Theta$ $\sim$ 19$^\circ$, which appears prominently in WAXS patterns of bulk PLLA \cite{PLLA_Miyata1997,PLLA_Zhang2005,PLLA_Kawai2007} and of PLLA--containing BCPs \cite{PLLA_Chen2006,PLLA_Kim1999}, is absent in all WAXS patterns displayed in Figure \ref{WAXS}.

\begin{figure}[H]
\centerline{\includegraphics[scale=0.6]{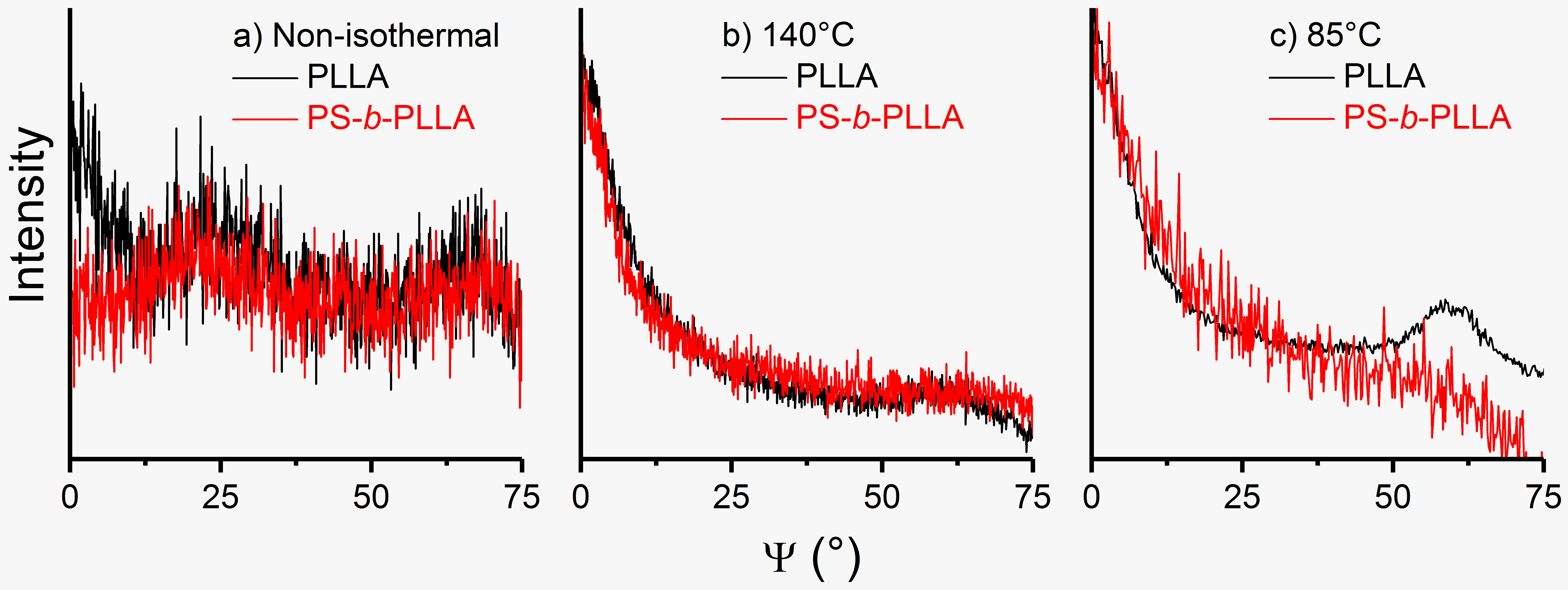}}
	\caption{Normalized Schulz scans of arrays of PS-\textit{b}-PLLA and PLLA nanorods located inside the aligned nanopores of self--ordered AAO (\textit{D} = 180 nm) for the (110)/(200) reflection of $\alpha$--PLLA a) after non--isothermal crystallization at --1 K/min; b) after isothermal crystallization at $T_C$ = 140$^\circ$C and c) after isothermal crystallization at $T_C$ = 85$^\circ$C. The intensity of all Schulz scans in panels b) and c) is normalized.}
	\label{psi}
\end{figure}

To evaluate the degree of crystal orientation, Schulz scans were carried out for the (110)/(200) composite reflection on PS-\textit{b}-PLLA and PLLA homopolymer crystallized non-isothermally at a cooling rate of --1 K/min (Figure \ref{psi}a), isothermally at $T_C$ = 140$^\circ$C (Figure \ref{psi}b) and isothermally at $T_C$ = 85$^\circ$C (Figure \ref{psi}c) inside AAO with a nanopore diameter of 180 nm. No indications of pronounced oriented crystallization were found for the non-isothermally crystallized samples (Figure \ref{psi}a). However, the Schulz scans of all isothermally crystallized samples show, independent of $T_C$, a maximum at $\Psi$ = 0$^\circ$. This maximum indicates the presence of a crystal population having the (110)/(200) lattice planes oriented parallel to the AAO membrane surfaces and normal to the AAO nanopore axes.  Moreover, the normalized Schulz scans obtained for PS-\textit{b}-PLLA and PLLA homopolymer isothermally crystallized at $T_C$ = 140$^\circ$C (Figure \ref{psi}b) and $T_C$ = 85$^\circ$C (Figure \ref{psi}c) inside AAO mostly coincide in the $\Psi$ range from 0$^\circ$ to $\sim$50$^\circ$ that comprises the peak at $\Psi$ = 0$^\circ$. The full width at half maximum of the peak at $\Psi$ = 0$^\circ$ lies for all Schulz scans showing such a peak between 7$^\circ$ and 9$^\circ$. Notably, the Schulz scan obtained for PLLA homopolymer crystallized at $T_C$ = 85$^\circ$C contains a second peak at $\Psi$ $\sim$ 60$^\circ$ (Figure \ref{psi}c). This maximum likely originates from the same crystal population as the maximum at $\Psi$ = 0$^\circ$ and reflects the pseudo-hexagonal structure of $\alpha$--PLLA. The same phenomenological crystal orientation was previously obtained for PLLA homopolymer in AAO by two-dimensional X--ray scattering \cite{PLLA_Guan2015}.

\section{Discussion}

\subsection{Melt morphologies}
It is reasonable to assume that the morphologies of the porous PS nanorods obtained via PLLA crystallization at $T_C$ = 85$^\circ$C (Figure \ref{TEM_85}) represent the nanoscopic domain structures which the PS-\textit{b}-PLLA melt forms inside the AAO nanopores. The following arguments corroborate this notion. 

1) The dominance of concentric cylinder shells oriented along the AAO nanopores for \textit{D}/\textit{L}$_0$ $\geq$ 4.8 (Figure \ref{TEM_85}a-c) is consistent with previous results obtained for fully amorphous symmetric BCPs under cylindrical confinement \cite{BCP-NP_Shin2004,BCP-NP_Xiang2004,BCP-NP_Xiang2005b,BCP-NP_Sun2005,BCP-NP_Ma2009}. 

2) BCPs are known to form helices and other morphology types not occurring in bulk systems when confined to rigid cylindrical nanopores. The occurrence of helical structure motifs inside cylindrical nanopores has been reported for non--crystallizable BCPs forming cylinders in the bulk \cite{BCP-NP_Xiang2005a,BCP-NP_Wang2009,BCP-NP_Dobriyal2009}. For example, in the 1.5 $\mu$m long fragment of a nanorod consisting of atactic asymmetric polystyrene-\textit{block}-poly(2-vinylpyridine) (PS-\textit{b}-P2VP) with PS as majority component \cite{BCP-NP_Wang2009} shown in Supporting Figure S4a the following structural motifs can be seen from the left to the right: a single P2VP helix winding about a straight central P2VP cylinder, straight P2VP cylinders aligned with the nanorod axis, a triple P2VP helix winding about a straight central P2VP cylinder and again a single P2VP helix winding about a straight central P2VP cylinder. A detail of a triple P2VP helix winding about a straight central P2VP cylinder is shown in Supporting Figure S4b. Complex morphologies based on helical structure motifs were also observed in silica nanorods prepared by BCP--templated sol-gel chemistry \cite{BCP-NP_Wu2004}. 

3) Both the nanorods consisting of symmetric PS-\textit{b}-PLLA (Figure \ref{TEM_85}a-d) and of asymmetric PS-\textit{b}-P2VP (Supporting Figure S4) consist of segments characterized by different types of internal morphologies; transitions between different morphology types occur within one and the same nanorod (Figure \ref{TEM_85}b--d and Supporting Figure S4a). Different morphology types occur within one PS-\textit{b}-PLLA nanorod even for \textit{D}/\textit{L}$_0$ values of $\approx$ 7.3, including complex structural motifs characterized by the presence of different types of intertwined and single helices (Figure \ref {TEM_85}d). This observation is in line with the theoretical prediction that several types of BCP melt morphologies with similar free energy ("degenerate structures") may co--exist within a rigid confining cylindrical nanopore \cite{BCP-NP_Yu2006}. Moreover, melt morphologies of BCPs in cylindrical confinement may represent kinetically trapped or even transient states rather than equilibrium morphologies \cite{BCP-NP_Sevink2001,BCP-NP_Sevink2008}. 

4) Even for metastable \textit{H}* phases of bulk asymmetric PS-\textit{b}-PLLA containing helical PLLA domains \cite{PLLA_Ho2004a,PLLA_Ho2009a,PLLA_Ho2009b} confined crystallization of PLLA was reported to occur \cite{PLLA_Chiang2005, PLLA_Chiang2009, PLLA_Zhao2013}. Moreover, thermal quenching of bulk melts of the symmetric PS-\textit{b}-PLLA used here results in the conservation of the ordered lamellar melt domain structure and in suppression of PLLA crystallization (Supporting Texts 1 and 2; Supporting Figures S1 and S3). Hence, it is plausible to assume that PLLA crystallization in PS-\textit{b}-PLLA inside AAO nanopores at $T_C$ = 85$^\circ$C was strictly confined by the nanoscopic PS-\textit{b}-PLLA melt domain structure fixated by the vitreous PS domains. 

Quenching melts of the symmetric PS-\textit{b}-PLLA used here confined to AAO followed by crystallization at $T_C$ = 85$^\circ$C reproducibly resulted in confined crystallization -- we did not find morphologies suggesting the occurrence of templated or breakout crystallization. For comparison, Supporting Figure S5 shows TEM images of PS nanorods prepared exactly in the same way as the PS nanorods displayed in Figure \ref{TEM_85}, the only difference being that a slightly asymmetric PS-\textit{b}-PLLA with PS as majority component was used. The PLLA formed lamellar crystals initially located at the positions of the voids in the PS nanorods, which were arranged in a fishbone-like way (side views in Supporting Figure S5a and b) and which had an ellipsoidal contour (top view in Supporting Figure S5d). It is obvious that in this case PLLA crystallization occurred in the breakout regime. Nevertheless, the X--ray scattering intensity profiles obtained by $\Theta$/2$\Theta$ scans (Supporting Figure S6) and Schulz scans (Supporting Figure S7) were very similar to the corresponding X--ray scattering intensity profiles obtained for symmetric PS-\textit{b}-PLLA inside AAO nanopores (Figures \ref{WAXS} and \ref{psi}). Even though the PLLA domains in the slightly asymmetric PS-\textit{b}-PLLA may exhibit better connectivity than in the symmetric PS-\textit{b}-PLLA (corresponding to the transition from lamellar to gyroidal bulk morphology), this outcome nevertheless indicates better compatibility of PLLA crystallization with the morphologies formed by the symmetric PS-\textit{b}-PLLA inside the AAO nanopores.       

\begin{figure}[H]
\centerline{\includegraphics[scale=0.08]{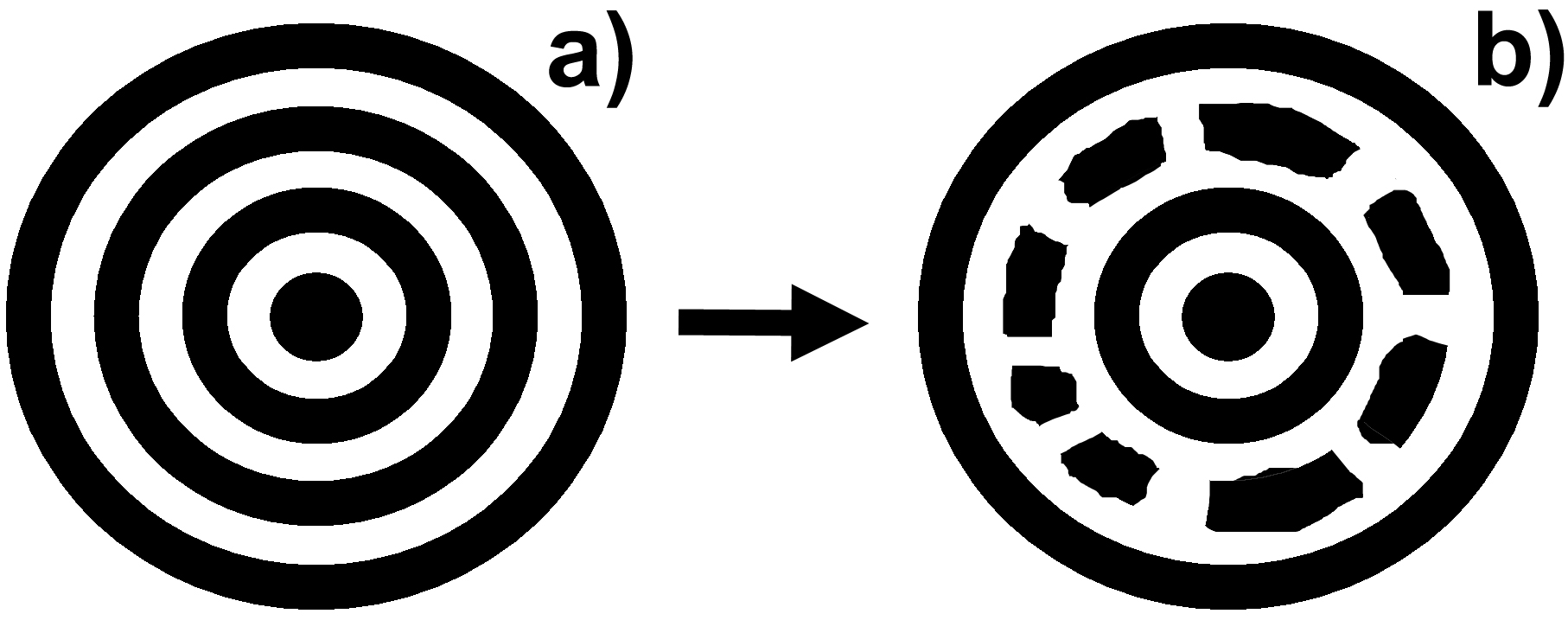}}
	\caption{Sketch of cross-sectional views of PS-\textit{b}-PLLA nanorods representing the structure evolution during annealing at 140$^\circ$C (PLLA black). a) Initially, undisturbed concentric cylinder shells exist. b) During annealing, the cylinder shells formed by the second-outermost PS-\textit{b}-PLLA layers are destroyed by an interplay of morphology ripening and PLLA crystallization.}
	\label{scheme_ripening}
\end{figure}

\subsection{Interplay of crystallization and morphology ripening}
We suggest that the morphologies of the PS nanorods obtained via isothermal crystallization of PS-\textit{b}-PLLA inside AAO at $T_C$ = 140$^\circ$C (Figures \ref{TEM_140} and \ref{TEM_310nm}) evolved as follows. After infiltration but prior to crystallization, the PS-\textit{b}-PLLA inside the AAO was subjected to extended annealing at 180$^\circ$C (cf. Experimental Section) during which the PS-\textit{b}-PLLA formed the morphologies represented by the PS nanorods displayed in Figure \ref{TEM_85}, namely intact concentric--cylinder--shell morphologies as dominating morphology type as well as morphologies comprising helical structure motifs. Subsequent annealing of the PS-\textit{b}-PLLA nanorods inside the AAO nanopores at $T_C$ = 140 $^\circ$ results in PLLA crystallization while PS and amorphous PLLA are still soft. Scrutinizing the morphologies of PS nanorods obtained by PLLA crystallization at $T_C$ = 140 $^\circ$ and subsequent removal of the PLLA reveals the following insights, as schematically displayed in Figure \ref{scheme_ripening}. 

1) The outermost PS cylinder shell remained mostly intact. This result indicates pronounced and persistent interface-induced ordering driven by PLLA adsorption on the AAO nanopore walls. The polyester PLLA likely segregates to the AAO nanopore walls since hydrogen bonds can be formed between the terminal hydroxyl groups at the AAO nanopore walls and the PLLA carbonyl groups. Moreover, polylactide homopolymer forms rigid amorphous fractions at inorganic oxidic surfaces \cite{PLLA_Klonos2016}. Therefore, the PLLA blocks forming the outermost concentric PLLA cylinder shell are likely pinned to the AAO nanopore walls. The concentric PS cylinder shell belonging to the outermost PS-\textit{b}-PLLA layer is encased by the outermost PLLA cylinder shell and contains many PS blocks covalently bond to PLLA blocks irreversibly adsorbed on the AAO nanopore walls. Hence, the concentric PS cylinder shell belonging to the outermost PS-\textit{b}-PLLA layer will be stabilized too so that it is neither affected by PLLA crystallization nor by ripening of the PS-\textit{b}-PLLA melt morphology. 

2) The second-outermost cylindrical PS-\textit{b}-PLLA layers consisting of the second-outermost PLLA and the second-outermost PS cylinder shells transformed into disturbed ring-like and helical structure motifs (Figures \ref{TEM_140} and \ref{TEM_310nm}), even though undisturbed concentric cylinder shells would have allowed unconfined growth of PLLA crystals along the AAO nanopores. This implies that at $T_C$ = 140$^\circ$ a mechanism for this transformation must exist that is neither effective above the PLLA melting point nor below the glass transition temperature of PS. Since the transformation of the second-outermost concentric PLLA and PS  cylinder shells impedes unconfined crystal growth along the second-outermost PLLA cylinder shell, it is reasonable to assume that growth of PLLA crystals is not the driving force for this morphological reconstruction. Instead, we assume that the second-outermost concentric PLLA and PS cylinder shells correspond to kinetically trapped non-equilibrium features formed by molten PS-\textit{b}-PLLA, which transform into more favorable morphologies while the PS-\textit{b}-PLLA subjected to this transformation is still soft and amorphous. It was predicted that symmetric BCPs under cylindrical confinement form concentric cylinder shells faster than competing morphologies \cite{BCP-NP_Sevink2001}. The infiltration of PS-\textit{b}-PLLA into the AAO nanopores driven by adhesion may further promote the formation of concentric-lamellar morphologies parallel to the flow direction. However, elastic frustrations of BCP molecules under cylindrical confinement may be better accommodated by helices and annuli having adaptable pitches and spacings \cite {BCP-NP_Sevink2008}. Such frustrations originating from the outermost PS-\textit{b}-PLLA layer probably forming rigid amorphous fractions may be transmitted to the second--outermost PS-\textit{b}-PLLA layer so that a conversion of the second-outermost concentric PLLA and PS cylinder shells into helical and/or annular structure motifs might be advantageous. The question arises as to why during extended annealing of PS-\textit{b}-PLLA confined to AAO above the melting point of PLLA (cf. Experimental Section) the concentric-lamellar morphology is persistent (so that it dominates in the PS-\textit{b}-PLLA nanorods crystallized at $T_C$ = 85$^\circ$C). We assume that the onset of PLLA crystallization in a temperature window below the melting temperature of PLLA where PS is still soft may catalyze the transformation of the second-outermost concentric PLLA and PS cylinder shells. Proceeding PLLA crystallization then freezes this process so that transient ripening states are arrested.

3) ``Low-pitch'' helices such as those shown in Figure \ref{TEM_140}b and c or such as the outermost helix in seen in Figure \ref{TEM_85}d with a pitch of a few 10 nm located next to the outermost intact PS layer were frequently found. This finding corroborates the notion that this helix type is advantageous -- it either survived heating to $T_C$ = 140 $^\circ$ or, alternatively, it might even be newly formed by the transformation of the second--outermost PS-\textit{b}-PLLA layers (see item 2 above).  

4) ``High--pitch'' helices with leads of a few 100 nm, as seen in Figure \ref{TEM_85}b-d, are another structure motif that had disappeared after annealing at $T_C$ = 140 $^\circ$. The ``high--pitch'' helices, which are located in the center of the PS-\textit{b}-PLLA nanorods, apparently tend to transform into straight concentric cylinder shells and straight central cylinders that allow unconfined growth of PLLA crystals along the AAO nanopores. This observation is in line with earlier reports by Chiang et al., who found that PLLA helices formed by asymmetric PS-\textit{b}-PLLA transform into straight PLLA cylinders if the PLLA is crystallized at a $T_C$ where PS is soft \cite{PLLA_Chiang2005,PLLA_Chiang2009}.

\subsection{Crystal growth}
While some authors identified both the (100) and (110) faces as preferred growth planes of $\alpha$--PLLA crystals \cite{PLLA_Abe2001, PLLA_Chiang2009}, other authors exclusively considered the (110) faces as preferential growth planes \cite{PLLA_Iwata1998}. Crystallization of PLLA homopolymer inside AAO nanopores results in preferential orientation of the (110)/(200) faces normal to the AAO nanopore axes, as reported by Guan et al. \cite{PLLA_Guan2015} and as evident from the isothermal crystallization experiments reported here (cf. Figures \ref{WAXS} and \ref{psi}). Hence, steady-state growth of PLLA homopolymer crystals along the AAO nanopore axes occurs normal to the (110)/(200) faces. The $\Theta$/2$\Theta$ patterns obtained for PS-\textit{b}-PLLA crystallized non--isothermally inside AAO nanopores at --1 K/min (Figure \ref{WAXS}) do not show the (110)/(200) reflection, which is typically the strongest $\alpha$--PLLA peak. This outcome indicates that the PLLA crystallites had not reached their steady-state growth shape. On the other hand, the occurrence of the (100) reflection is rather unusual. Thus, non-isothermal crystallization at --1 K/min of PS-\textit{b}-PLLA inside the AAO nanopores indicates that at early crystallization stages the (100) face contributes to crystal growth while the contribution of the (110) face is negligible. 

Independent of $T_C$, isothermal crystallization inside AAO nanopores of both PS-\textit{b}-PLLA and PLLA homopolymer resulted in preferred orientation of the (110)/(200) faces of $\alpha$--PLLA normal to the AAO nanopore axes (Figures \ref{WAXS} and \ref{psi}). PLLA crystals with this preferred orientation have, during their steady-state growth, their fastest growth direction oriented parallel to the AAO nanopore axes. If the PS-\textit{b}-PLLA nanorods inside the AAO nanopores exhibit concentric--cylinder--shell morphologies, these PLLA crystals can rapidly grow along the AAO nanopores within concentric cylinder shells so that they prevail over PLLA crystals with other orientations. However, it is noticeable that PS-\textit{b}-PLLA inside AAO nanopores exhibits the same kind and the same degree of phenomenological PLLA crystal orientation as PLLA homopolymer. The PS-\textit{b}-PLLA nanorods inside the AAO nanopores contain morphological features such as helices and the PLLA domains in the second-outermost PS-\textit{b}-PLLA layer formed at $T_C$ = 140$^\circ$C that can be considered as obstacles to unconfined PLLA crystal growth along the AAO nanopores. Chiang et al. found that in bulk asymmetric PS-\textit{b}-PLLA forming metastable $H$* phases comprising helical PLLA domains the direction of fastest crystal growth confined by vitreous PS is aligned with the straight central axis rather than with the curvilinear helical track of the helical PLLA domains \cite{PLLA_Chiang2009}. The results obtained here also suggest that locally occurring tortuous growth paths for the PLLA crystals within PS-\textit{b}-PLLA nanorods do not reduce the degree of crystal orientation. 

\section{Conclusions}
We have investigated the crystallization of the PLLA blocks of lamellae-forming PS-\textit{b}-PLLA inside straight cylindrical nanopores of AAO under weak confinement ($D$/$L_0 \geq$ 4.8). The melt morphology predominantly consists of concentric cylinder shells oriented parallel to the AAO nanopores. Even for $D$/$L_0 \approx$ 7.3 the PS-\textit{b}-PLLA nanorods contain helices with pitches of the order of $L_0$ and intertwined multiple helices with leads several times larger than $L_0$. Within a nanorod transitions between different morphology types may occur. If the PS-\textit{b}-PLLA inside the AAO nanopores is quenched to $T_C$ = 85$^\circ$C, PLLA crystallization is strictly confined by the vitrified PS domains. At $T_C$ = 140$^\circ$C (PS is soft) the outermost PS-\textit{b}-PLLA layer remains intact due to irreversible adsorption at the AAO nanopore walls. The onset of PLLA crystallization apparently triggers the transformation of the concentric PLLA and PS cylinder shells initially forming the second-outermost PS-\textit{b}-PLLA layer into distorted helical or annular structure motifs to mitigate elastic frustrations transmitted from the immobilized outermost PS-\textit{b}-PLLA layer. Proceeding PLLA crystallization then arrests transient states of this morphological reconstruction. Helices with pitches of the order of $L_0$ formed by the second--outermost PS-\textit{b}-PLLA layer as well as concentric cylinder shells and central cylindrical domains closer to the center of the PS-\textit{b}-PLLA nanorods persist extended annealing at $T_C$ = 140$^\circ$C. Moreover, PLLA crystallization drives the conversion of intertwined helix systems with leads of a few 100 nm located in the center of the PS-\textit{b}-PLLA nanorods into straight concentric cylinder shells or straight central cylinders. The direction of fastest PLLA crystal growth is, independent of $T_C$, aligned with the AAO nanopore axes to the same degree as for PLLA homopolymer. The hierarchical confinement of AAO nanopores and BCP melt domain structure reduces the impact of heterogeneous nucleation and retards crystallization of the PLLA. Thus, early stages of polymer crystallization preceding the stationary crystal growth state may be captured \textit{via} the hierarchical confinement of semicrystalline BCPs in AAO nanopores.

\begin{acknowledgement}
The authors thank the European Research Council (ERC-CoG-2014, project 646742 INCANA; Marie Sklodowska-Curie grant 706329/cOMPoSe) for funding. Technical support by K. Sklarek (MPI for Microstructure Physics, Halle), by C. Hess and H. Tobergte (Universit\"at Osnabr\"uck) as well as by K. Herfurt (Martin-Luther-Universit{\"a}t Halle-Wittenberg) is gratefully acknowledged.
\end{acknowledgement}

\bibliography{PLLA_and_PS-b-PLLA,Crystallization_in_nanopores,BCPs_in_nanopores}

\newpage
\center{\textbf{Semicrystalline block copolymers in rigid confining nanopores}}
\\
\center{Man Yan Eric Yau, Ilja Gunkel, Brigitte Hartmann-Azanza, Wajiha Akram, Yong Wang, Thomas Thurn-Albrecht, Martin Steinhart}

\vspace*{1cm}

\begin{figure}[H]
\centerline{\includegraphics[scale=1]{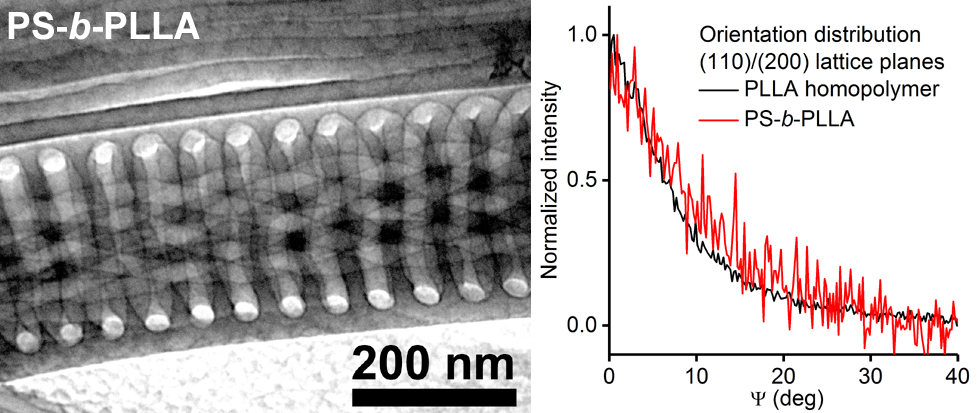}}
\end{figure}
\center{Table of Contents Figure}
\end{document}